\documentclass[aps,pre,twocolumn,showpacs]{revtex4-2}
\usepackage{amsmath,amssymb,graphicx}
\usepackage{ragged2e}
\usepackage{epsfig}
\usepackage{graphicx}
\usepackage{overpic}
\usepackage{textcomp}
\usepackage[titletoc,title]{appendix}
\usepackage{soul}
\usepackage{graphicx}
\usepackage{placeins}
\usepackage{subcaption}
\usepackage[caption=false,font=footnotesize,captionskip=-3mm,farskip=-0mm]{subfig}
\usepackage{color}

\def\black{\textcolor{black}}

\begin{document}

\title{Short-time statistics of extinction and blowup in reaction kinetics}

\author{Rotem Degany}
\author{Michael Assaf}
\author{Baruch Meerson}
\affiliation{Racah Institute of Physics, Hebrew University of Jerusalem, Jerusalem 91904, Israel}

\begin{abstract}
We study the statistics of extinction and blowup times in well-mixed systems of stochastically reacting particles. We focus on the short-time tail, $T \to 0$, of the extinction- or blowup-time distribution $\mathcal{P}_m(T)$, where $m$ is the number of particles at $t=0$. This tail often exhibits an essential singularity at $T=0$, and we show that the singularity is captured by a time-dependent WKB (Wentzel–Kramers–Brillouin) approximation applied directly to the master equation. This approximation, however, leaves undetermined a large pre-exponential factor. 
We show how to calculate this factor by applying a leading- and a subleading-order WKB approximation to the Laplace-transformed backward master equation. Accurate asymptotic results can be obtained when this WKB solution can be matched to  another approximate solution (the ``inner" solution), valid for not too large $m$. We demonstrate and verify this method on three examples of reactions which are also solvable without approximations.

\end{abstract}

\maketitle
\section{Introduction}
This paper deals with systems of stochastically reacting particles which exhibit one of two extreme forms of first-passage behavior to an absorbing state: a finite-time population extinction, when the number of particles which initially was large hits zero, or a finite-time population blowup, when the number of particles which initially was small becomes infinite. Population extinction  is encountered in fields ranging from chemical physics and population biology to ecology and epidemiology \cite{Delbruck,Bartlett,Karlin,Bailey,OvaskainenM,Allen}. In its turn, population blowup is a useful simplification of a rapid (super-Malthusian) population proliferation \cite{Johanssen}.
Because of the intrinsic randomness of the elemental processes, the extinction (or blowup) time $T$ in these systems is a random quantity, and it is interesting and sometimes crucial to determine its probability distribution $\mathcal{P}(T)$.

For simplicity we assume a well-mixed system of single-species particles (we call them $A$) undergoing Markovian stochastic combinatorial reactions:  jump processes such as $A\!\to\! 0$,  $2A \!\to\! 0$, $2A \!\to\! A$, $2A\!\to\! 3A$, etc. These reactions can  be described by the master equation for the probability $P_n(t)$ of observing $n$ particles at time $t$. In its turn, the extinction- or blowup-time probability distribution $\mathcal{P}_m(T)$, where $m$ is the initial number of particles in the system, is governed by the probability flux to the absorbing state of zero or infinite population size, respectively. 

Here we are interested in \emph{unusually fast} extinctions or blowups. Such rare large deviations are described by the short-time tail, $T \to 0$, of the extinction or blowup time probability distribution $\mathcal{P}_m(T)$ or, equivalently, by the large-$s$ asymptotic of the Laplace transform of $\mathcal{P}_m(T)$,
\begin{equation}
\label{LT}
R(s,m)=\int_0^\infty dT\,e^{-sT} \mathcal{P}_m(T)\,.
\end{equation}
For a whole class of stochastic reaction systems, the $\mathcal{P}_m(T\to 0)$ tail shows a nontrivial behavior. In particular, it exhibits an essential singularity at $T=0$; that is, it vanishes at $T=0$ together with all its derivatives with respect to $T$. 

The  probability distribution $\mathcal{P}_m(T)$ can be represented in terms of a superposition of eigenmodes of the generator of the master equation, see e.g. Refs.~\cite{AM2010,assaf2017wkb}.   Even in simple cases where the eigenmodes, and the corresponding eigenvalues, can be determined in analytical form, the resulting sum over the eigenmodes is not very informative for the description of the $\mathcal{P}_m(T\to 0)$ tail. 

A simple alternative to the spectral language is provided by the time-dependent WKB (Wentzel-Kramers-Brillouin) theory, directly applied to the master equation of the system via the WKB ansatz $P_n(t) = \exp[-S(n,t)]$ \cite{Elgart_Kamenev_2004}. \black{This variant of WKB method relies on the small parameters $T/\bar{T}\ll 1$ (where 
$\bar{T}$ is the mean time to extinction or blowup, respectively) and $1/m\ll 1$}. This ansatz, and the subsequent approximation,  effectively replace the original master equation by a Hamilton-Jacobi equation for the action $S(n,t)$, \black{where $n$ plays the role of coordinate}. The latter can be solved for the \emph{optimal} \black{---that is, most likely---} path $n(t)$ of the conditioned process, which gives a dominant contribution to the  $\mathcal{P}(T\to 0)$ tail. The optimal path obeys the initial condition $n(t=0)=m$  and the final condition $n(T)=0$ or $n(T)=\infty$ for the extinction or blowup, respectively. For a well-mixed single population of particles with time-independent process rates, the ensuing effective classical mechanics problem is always integrable.

The leading-order time-dependent WKB theory is quite informative. Most importantly, it correctly describes the leading-order, exponential behavior of $\mathcal{P}(T\to 0)$, including an essential singularity at $T=0$.  It also gives the optimal path of the conditioned process which is interesting in itself.  This theory misses, however,  pre-exponential factors which can be very large, hence important. To account for these, one could go to the next, subleading WKB order. In practice this step is technically involved because of the explicit time dependence of this variant of WKB theory. 

As we show here, a much better alternative is to work with the Laplace-transformed backward master equation for the function $R(s,m)$ defined in Eq.~(\ref{LT}). \black{The backward master equation describes the time evolution of probabilities  by integrating from a future state backward in time. This equation is convenient for an explicit tracking of the dependence of the solution on the initial condition (see e.g. Refs. \cite{gardiner,Redner_2001,Krapivsky_Redner_Ben-Naim_2010} for detailed explanations and derivations), and it is often used for the analysis of first-passage problems. Crucially, upon the Laplace transform in time, the backward master equation becomes time-independent}, which greatly simplifies the analysis. Here we show that, using the Laplace parameter $s$ as the large parameter (which is equivalent to \black{the same small parameter $T/\bar{T}\ll 1$ as in the time-dependent WKB method)} alongside the strong inequality $m\gg 1$ \black{(a large number of particles at $t=0$)}, one can develop a leading and subleading WKB approximation in a straightforward manner.  For continuous Markovian stochastic processes, described by a Langevin equation with additive white Gaussian noise, a similar method  has been recently developed in Ref. \cite{M2025b}.

As we will see shortly, in the extinction case this variant of WKB approximation captures both the leading-order exponential dependence of 
$\mathcal{P}(T)$, and the large $T$-dependent subleading prefactor, missing only an 
 $O(1)$ factor. The latter can also be determined if one can find an additional approximate solution to the Laplace-transformed backward master equation in the ``inner" region, that is for not too large $m$, including the region of $m=O(1)$ where the WKB solution is invalid. The lacking factor $O(1)$ can then be determined by matching the inner solution with the (leading and subleading) WKB solution in their common region of validity.  In the blowup case, when starting from a small number of particles, the matching with the inner solution is necessary for determining the entire subleading prefactor. 

Rather than presenting the method in full generality, we demonstrate it by considering three examples which cover both extinction and blowup phenomena. These examples are also solvable exactly, which allows us to verify the method in all details.

Here is a plan for the remainder of the paper. In Sec.~\ref{Annihilation} we consider the well-studied problem of population extinction resulting from the annihilation reaction $2A \to 0$. We start  Sec.~\ref{Annihilation}  with the (leading order) time-dependent WKB theory which  yields the $\mathcal{P}(T\to 0)$ tail, but only up to a preexponential factor.
We then use the Laplace-transformed backward master equation, and determine the preexponential factor by
(i) applying the leading- and subleading WKB method,  (ii) obtaining the inner solution, and (iii) matching the two solutions in their common validity region.  In Sec.~\ref{coalescence_and_death} we determine the $\mathcal{P}(T\to 0)$ tail in the system of two reactions leading to  population extinction: the coalescence $2A \to A$ and decay $A\to 0$.  In Sec.~\ref{superMalthus} we consider the binary branching reaction $2A \to 3A$ which leads to a finite-time blowup---a super-Malthusian catastrophe. 
Each of the three examples, presented in Sections~\ref{Annihilation}-\ref{superMalthus}, can be also solved exactly, which allows us to verify our asymptotics of $\mathcal{P}(T\to 0)$ by comparing them with those extracted from the exact solutions. 

Our main results are summarized in Sec.~\ref{discussion}.
For completeness, the Appendix presents the leading-order time-dependent WKB solution for the model in Sec.~\ref{coalescence_and_death}.

\section{Annihilation}
\label{Annihilation}
Our first example is the well-studied  binary annihilation reaction $2A \stackrel{\lambda}{\to} 0$. When starting from an even number of particles $m$, a well mixed population of particles undergoing this reaction ultimately goes extinct with probability $1$. The probability distribution $\mathcal{P}(T)$ of the extinction time in this system was determined exactly, in terms of a spectral decomposition, as early as in 1964 \cite{McQuarrie}.  The mean time to extinction (MTE) in this system is $\bar{T}=O(1/\lambda)$, and extinction times much shorter than $\bar{T}$ are large deviations which are the focus of our interest. 

The evolution of the probability $P_n(t)$ of observing $n$ particles at time $t$ in this system is described by the \black{forward} master equation~\cite{McQuarrie,AssafMeerson2006} 
\begin{equation}\label{mastereq}
\dot{P}_n(t) = r_{n+2} P_{n+2}(t)-r_n P_n(t),
\end{equation}
where $r_n=\frac{1}{2}n(n-1)$, and time is rescaled so that $\lambda=1$. 

The backward master equation, governing the probability distribution of extinction occurring at time $T$ given an initial population of $m$ particles (we assume for simplicity that $m$ is an even number),  reads 
\begin{equation}
\label{annih1}
\partial_T\mathcal{P}_m(T)=r_m[\mathcal{P}_{m-2}(T)-\mathcal{P}_m(T)]\,.
\end{equation} 
The boundary condition is $\mathcal{P}_{m=0}(T)=\delta(T)$, which corresponds to the fact that for an initial population of zero particles extinction occurs immediately.  Before we deal with $\mathcal{P}_m(T)$, let us calculate
the MTE, $\bar{T}_m$. One can do so by solving  a recursive equation, related to the backward master equation \cite{gardiner}:
\begin{equation}
\label{avtimeannih}
\bar{T}_{m}=\bar{T}_{m-2} + 1/r_m\,.
\end{equation}
The obvious boundary condition is $\bar{T}_{m=0}=0$, and the solution is
\begin{equation}
\bar{T}_m=2 \left(H_m-H_{m/2}\right)\,,\quad m=2,4,6,\dots\,.
\end{equation}
where $H_k$ is the harmonic number. For example, $\bar{T}_2=1$, $\bar{T}_4=7/6$, $\bar{T}_6=37/30$, etc. Notably,  the MTE remains finite in the limit of $m\to\infty$: $\bar{T}_{m\to \infty}= \ln 4$.

\black{\subsection{$\mathcal{P}_m(T)$: Exact Result and Asymptotics}}
To calculate $\mathcal{P}_m(T)$, we first solve the backward master equation (\ref{annih1}) in Laplace space:
\begin{eqnarray} \label{annihilation_laplace}
    sR(s,m)=r_m[R(s,m-2)-R(s,m)]\,,
\end{eqnarray}
with the transformed boundary condition $R(s,m=0)=1$. \black{Here $s$ is the Laplace conjugate of the extinction time $T$, see Eq.~(\ref{LT}).} The solution is
\begin{equation}\label{Rm}
R(s,m) = \frac{2^{-m} m! \,}{ \left(\frac{3-\sqrt{1-8 s}}{4}\right)_{m/2}  \left(\frac{3+\sqrt{1-8 s}}{4}\right)_{m/2}  }\,,
\end{equation}
where $(a)_k \equiv \Gamma(a+k)/\Gamma(a)$ is the Pochhammer symbol, and we continue to assume that $m$ is an even number. In particular, we find
\begin{eqnarray}
  R(s,2) &=& \frac{1}{s+1}\,, \nonumber\\
  R(s,4) &=& \frac{6}{(s+1) (s+6)}\,, \nonumber\\
  R(s,6) &=& \frac{90}{(s+1) (s+6) (s+15)}\,, \nonumber\\
  R(s,8) &=& \frac{2520}{(s+1) (s+6) (s+15) (s+28)}\,.\label{Rm2to8}
\end{eqnarray}
In general, an increase of $m$ by $2$ adds one pole to $R(s,m)$,  and each pole corresponds to an eigenmode of the generator of the stochastic process.  A convenient benchmark is provided by the limit $m\to \infty$, where the solution $R(s,m\to \infty) \equiv R(s)$ of Eq.~(\ref{Rm}) takes a universal form:
\begin{equation}\label{Runiv}
R(s) =\frac{\Gamma \left(\frac{3}{4}-\frac{1}{4} \sqrt{1-8 s}\right) \Gamma \left(\frac{3}{4}+\frac{1}{4} \sqrt{1-8 s}\right)}{\sqrt{\pi }}\,,
\end{equation}
where $\Gamma(\dots)$ is the gamma function. The function $R(s)$ has an infinite number of poles, corresponding to an infinite series in the spectral decomposition of $P_n(t)$ \cite{McQuarrie,AssafMeerson2006}.  \black{(A pole arises whenever the argument of one of the two Gamma functions becomes a negative integer $z=0,-1,-2, ...$, which occurs for $s=-1+3z-2z^2 = -1,-6,-15, ...$.)}

The pole representation is especially convenient for extracting the \emph{large}-$T$ tail of the distribution,
\begin{equation}\label{longtime}
\mathcal{P}(T \gg \bar{T}) \simeq \frac{3}{2}\, e^{-T}\,,
\end{equation}
because this tail is determined by a single pole of $R(s)$: the one closest to the origin, which is $s=-1$.

\begin{figure}[ht]
\includegraphics[width=2.7 in,clip=]{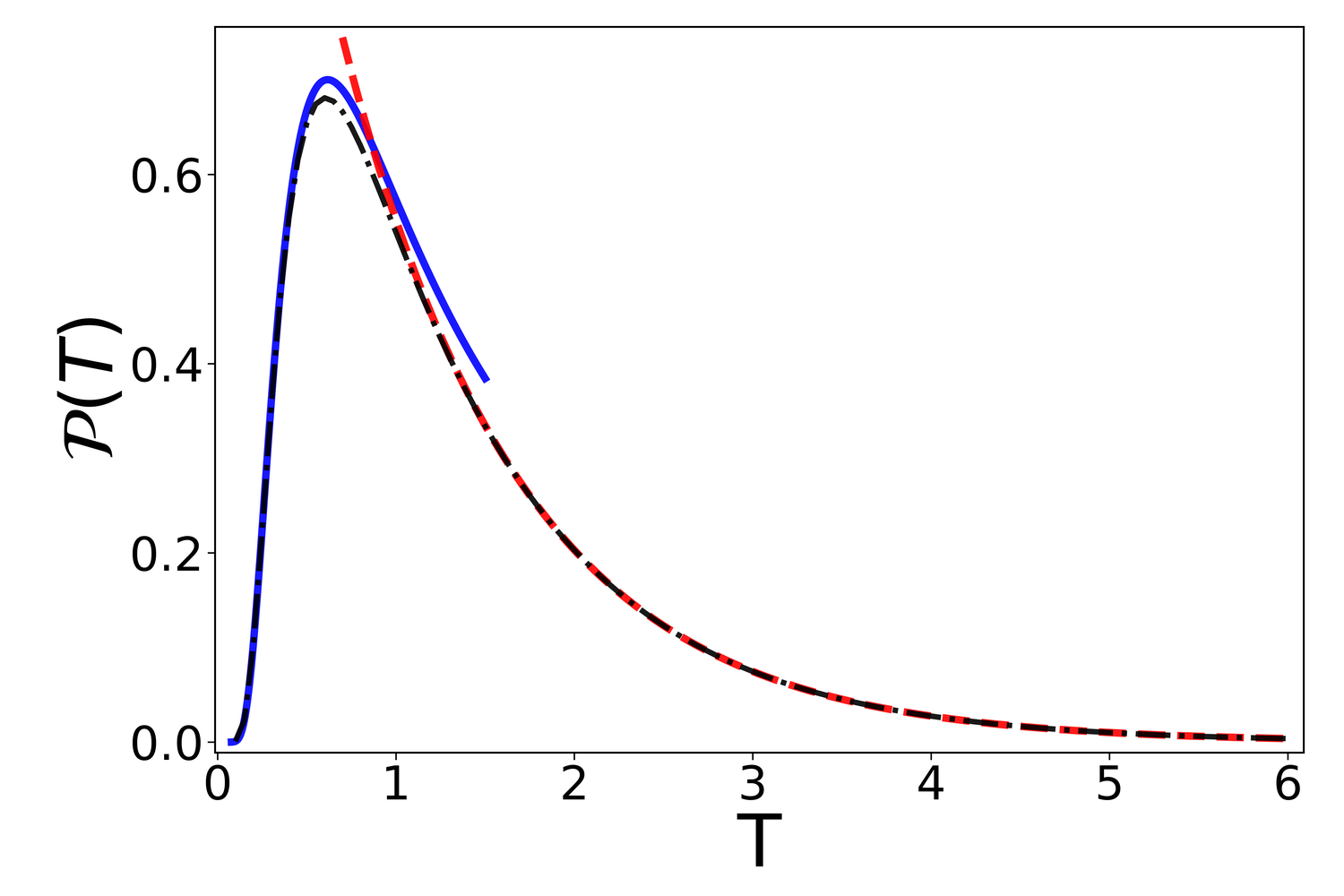}
\vspace{-4mm}\caption{ \justifying {The universal ($m\to \infty$) annihilation time probability distribution $\mathcal{P}(T)$ (black dotted-dashed line), the long-time tail~(\ref{longtime}) (red dashed line) and the short-time tail~(\ref{shorttime}) (blue solid line). }}
\label{fig1}
\end{figure}

The small-$T$ tail, which we are mostly interested in, is governed by \emph{all} the poles of $R(s)$. It is determined by the large-$s$ asymptotic of $R(s)$, which can be obtained by using \black{Stirling's approximation of the gamma function in Eq.~(\ref{Runiv}):}
\begin{equation}\label{Runivlarges}
R(s\to \infty)\simeq 2^{3/4} \sqrt{\pi } s^{1/4} e^{-\frac{\pi  \sqrt{s}}{\sqrt{2}}}\,.
\end{equation}
Its inverse Laplace transform (see, e.g. Ref. \cite{bateman1954tables}) immediately gives the desired small-$T$ tail:
\begin{equation}
\mathcal{P}(T \to 0)\simeq \frac{\pi ^{3/2}}{2 \sqrt{2} T^2}\, e^{-\frac{\pi ^2}{8 T}}\,.
\label{shorttime}
\end{equation} 
As one can see, this  tail exhibits an essential singularity at $T=0$, and it also has a large $O(T^{-2})$ prefactor.

Figure~\ref{fig1} shows the universal extinction time probability distribution $\mathcal{P}(T)$, obtained  by a numerical inverse Laplace transform of $R(s)$ from Eq.~(\ref{Runiv}). Also shown are the large- and small-$T$ asymptotics of  $\mathcal{P}(T)$: Eqs.~(\ref{longtime}) and (\ref{shorttime}), respectively. 

\black{\subsection{Time-Dependent WKB Approximation}}
Until now we have been discussing the exact results for $\mathcal{P}(T)$ and asymptotics extracted from these exact results. Now we will show that the small-$T$ tail of $\mathcal{P}(T)$ behavior can be captured -- but only up to the pre-exponential factor $O(T^{-2})$ -- by the time-dependent WKB approach \cite{Elgart_Kamenev_2004}.  To this end we apply the leading-order time-dependent WKB ansatz $P_n(t) = \exp[-S(n,t)]$ directly to the master equation (\ref{mastereq}) and assume that $n\gg 1$, so that
$ S(n+2,t) \simeq  S(n,t)+2 S'(n,t)$, where the prime denotes the derivative with respect to $n$. This approximation leads to the Hamilton-Jacobi equation
\begin{equation}\label{HJeq}
\partial_t S+H(n,\partial_n S) = 0\,,
\end{equation}
with the effective Hamiltonian
\begin{equation}\label{H}
H(n,p) = \frac{n^2}{2}\left(e^{-2p}-1\right)\,,
\end{equation}
which is an integral of motion:  $H(n,p)=E =\text{const}$. The Hamilton equations are
\begin{eqnarray}
  \dot{n} = -n^2 e^{-2p},\quad\quad\dot{p} = n \left(1-e^{-2p}\right)\,,\label{neqannihilation}
\end{eqnarray}
where the dots denote the time derivatives. With the integral of motion $H(n,p)=E$, the equation for $\dot{n}$ becomes

\begin{equation}\label{neq1}
\dot{n} = -n^2-2E\,,
\end{equation}
Equation~(\ref{neq1})  allows one to determine the time to extinction $T$ as a function of $E$ and the number of particles at $t=0$. In the benchmark case of infinitely many particles at $t=0$ we obtain
\begin{equation}\label{exttime}
T(E) = -\int_{\infty}^0 \frac{dn}{n^2+2E}= \frac{\pi}{\sqrt{8E}}\,.
\end{equation}
Inverting this relation, we obtain $E=\pi^2/(8T^2)$. In its turn, the extinction action is
\begin{equation}\label{S0}
S_0=\int_{\infty}^0 p(n,E) dn - E T = \pi \sqrt{\frac{E}{8}}=\frac{\pi^2}{8T}\,,
\end{equation}
leading to the extinction time probability distribution 
\begin{equation}\label{P(T)lead}
\mathcal{P}(T) \sim e^{-\frac{\pi^2}{{8T}}}.
\end{equation}
The applicability condition of this asymptotic is $T\ll \bar{T}=O(1)$. Comparing Eq.~(\ref{P(T)lead}) with Eq.~(\ref{shorttime}), we see that Eq.~(\ref{P(T)lead}) captures  the most important leading-order dependence on $T$ which describes the essential singularity at $T\to0$. It misses, however, the large pre-exponential factor $\sim T^{-2}\gg 1$. 

For the optimal path itself we obtain
\begin{equation}\label{n(t)}
n(t) = \frac{\pi}{2T}\cot \left(\frac{\pi  t}{2 T}\right)\,.
\end{equation}
 At $t\ll T$  it behaves as $n(t)\simeq 1/t$, which  coincides with the solution of the deterministic equation $\dot{n}(t)=-n^2(t)$ for the initial condition $n(t=0) = \infty$. This is to be expected as, at very short times, when the population size is still very large, the fluctuations are not needed: $p(t)=0$.  (Note that a purely deterministic solution for $n(t)$ would never reach $n=0$ in a finite time.)

This example clearly shows the advantages of the time-dependent WKB method:  it  provides the optimal path of the process,  captures the leading-order behavior of the $\mathcal{P}(T\ll \bar{T})$ tail, and is easily implementable.  Its disadvantage is that it misses the important (large) prefactor. 

 \black{\subsection{Laplace-Space WKB Approximation}}
We will now show how to reproduce the latter by employing the leading and subleading WKB formalism in  the Laplace space. Thus, we return to Eq.~(\ref{annihilation_laplace}) and pretend that we do not know its exact solution. We make the WKB ansatz  
\begin{equation}\label{WKBansatz}
R_{\text{WKB}}(s,m) = e^{-S(s,m)}=e^{-S_0(s,m)-S_1(s,m)}\,,
\end{equation}
assume that $S_0 \gg S_1$ and exploit the large parameters $s\gg 1$ (short times) and $m\gg 1$ (large population size).  As a result, $S_0$ and $S_1$ can be expanded as 
$$ 
S_0(s,m-2) \simeq S_0(s,m)-2 S_0'(s,m)+2S_0''(s,m)
$$ and 
$$
S_1(s,m-2) \simeq S_1(s,m)-2 S_1'(s,m)\,,
$$
where the primes denote derivatives with respect to $m$. Applying these to Eq.~(\ref{annihilation_laplace}) we obtain in the leading and sub-leading order: 
\begin{eqnarray}
    S_0'(s,m)&=&\frac{1}{2}\ln{\left(1+\frac{2s}{m^2}\right)},\nonumber \\
    S_1'(s,m)&=&-\frac{s}{m^3+2 m s} \label{annihilation_sublead}.
\end{eqnarray}

In the leading order we can integrate the equation for $S_0'(s,m)$ from $0$ to $m$ and use the condition $S_0(m=0)=0$ which follows from the boundary condition $R(m=0,s)=1$. 
In the subleading order we need to account for the fact that the WKB approximation becomes invalid  at small $m$. Therefore we integrate the equation for $S_1'(s,m)$ from an a priori unknown cutoff $m_0=O(1)$ to $m$. We obtain
\begin{eqnarray}
  &&S_0(s,m) =  \frac{m}{2} \ln \left(1+\frac{2
   s}{m^2}\right)+\sqrt{2s} 
   \arctan\left(\frac{m}{\sqrt{2s}}\right)\,,
   \label{S0forR} \nonumber \\
&&   S_1(s,m)=\frac{1}{4}\ln
   \left(\frac{1+2s/m^2}{1+2s/m_0^2}\right)\,.
\end{eqnarray}
This procedure, which employs two large parameters, $s\gg 1$ and $m\gg 1$,  yields the WKB solution via Eq.~(\ref{WKBansatz}) up to the unknown factor $m_0$.  To determine this factor, we will now solve Eq.~(\ref{annihilation_laplace}) in the ``inner" region $m \ll \sqrt{s}$ and match the inner solution to the $m \ll \sqrt{s}$ asymptotic of the WKB solution. 

\begin{figure}[ht]
\includegraphics[width=2.7 in,clip=]{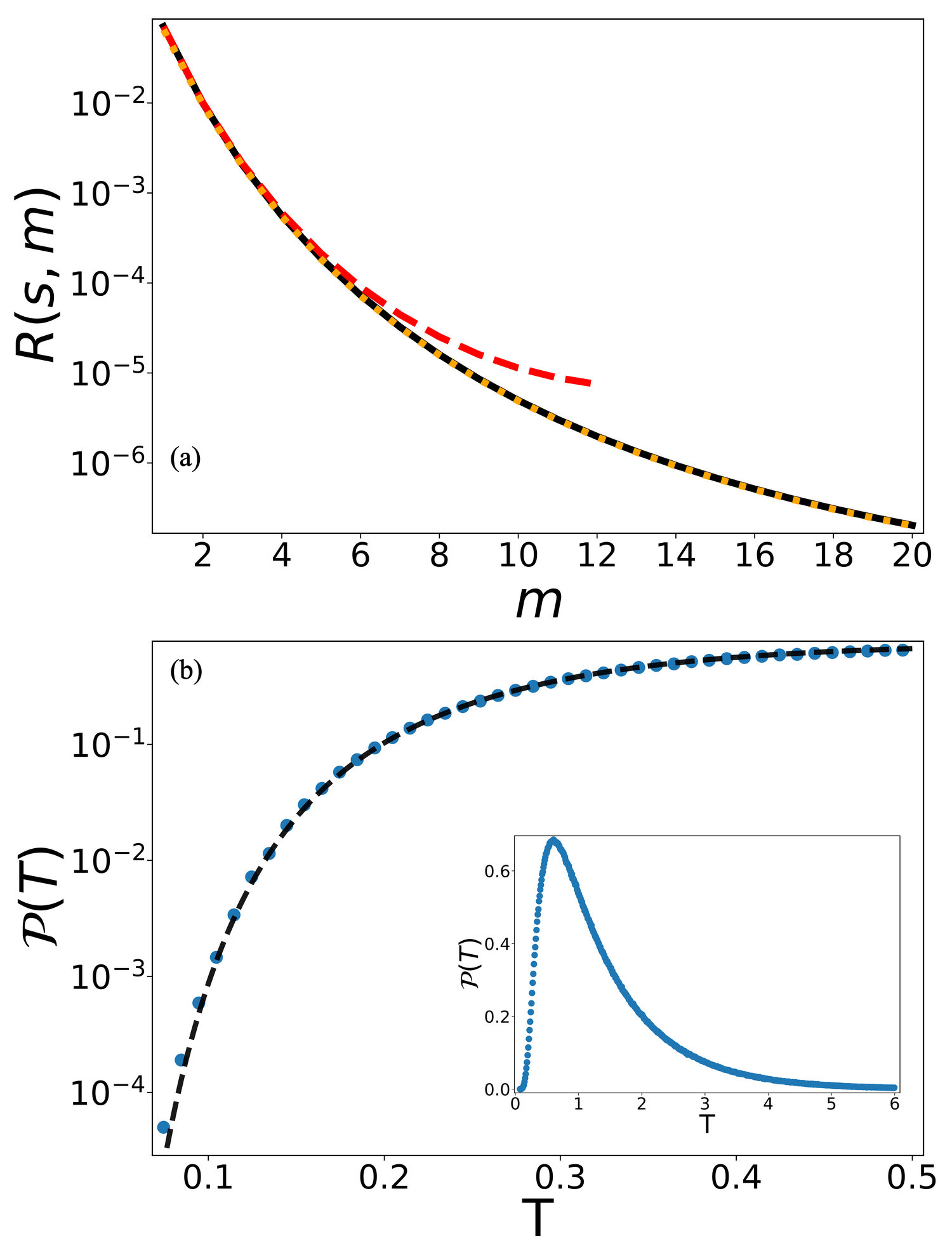}
\vspace{-4mm}\caption{\justifying {(a) Solutions  of Eq.~(\ref{annihilation_laplace}) for the annihilation: exact (black solid line), inner (red dashed line), and the leading and subleading WKB  (orange dotted line), for $s=10^2$. (b) Probability distribution $\mathcal{P}(T)$ of extinction times. The short-time distribution tail (\ref{shorttime}) (black dashed line) is compared with numerical simulations (blue dots), with $10^7$ realizations. The inset shows the simulation results for $\mathcal{P}(T)$ (in linear scale) over a longer time scale. } }
\label{fig2}
\end{figure}

In the inner region  Eq.~(\ref{annihilation_laplace}) gets simplified, as one can neglect 
the term $r_m R(s,m)$ compared with the two other terms of the equation. \black{This can be seen by rearranging Eq.~(\ref{annihilation_laplace}) as}
$$
R(s,m)=\frac{r_m}{r_m+s}R(s,m-2)
$$ 
\black{and neglecting $r_m$ compared with $s\gg 1$ in the denominator.}
The resulting approximate equation
\begin{equation}
r_m R(s,m-2)=sR(s,m)\,,
\end{equation}
with the boundary condition $R(m=0)=1$, can be easily solved. The solution is
\begin{equation}
\label{innerann}
R_{<}(s,m) = (2s)^{-m/2} m!\,, \quad 0<m \ll \sqrt{s}.
\end{equation}
As an illustration of the validity of the inner solution, Fig.~\ref{fig2} compares the exact $R(s,m)$ and the inner solution~(\ref{innerann}) for $s=10^2$. As one can see, the agreement is very good as long as $m$ is not too large.

In its turn, the $m\ll \sqrt{s}$ asymptotic of $R_{\text{WKB}}$~(\ref{WKBansatz}) is 
\begin{equation}
\label{asympWKBann}
R_{\text{WKB}}(s,m\ll \sqrt{s})\simeq \sqrt{\frac{m}{m_0}} \left(\frac{m^2}{2s}\right)^m e^{-m}\,.
\end{equation}
Matching the two asymptotics (\ref{innerann}) and (\ref{asympWKBann}) in their common validity region 
$1\ll m\ll \sqrt{s}$ and using Stirling's formula for $m!$, we obtain $m_0 = 1/(2\pi)$.  
Now the WKB expression, Eq.~(\ref{WKBansatz}), determines $R(s)$ completely.  

The expressions for $S_0(s,m)$ and $S_1(s,m)$ simplify in the universal limit $m\to \infty$:
\begin{equation} \label{Annihilation_Action}
S_0(s)=\pi\sqrt{\frac{s}{2}}, \quad S_1(s)=-\frac{1}{4}\,\ln\,(8\pi^2s)\,.  
\end{equation}
The resulting $R(s) = \exp[-S_0(s)-S_1(s)]$ coincides with Eq.~(\ref{Runivlarges}) leading to the $\mathcal{P}(T\to 0)$ tail described by Eq.~(\ref{shorttime}), including all preexponential factors. 

Figure~\ref{fig2}(a) compares the (leading and subleading) WKB asymptotic~(\ref{WKBansatz}) and inner solution (\ref{innerann}) for $R(s,m)$ as functions of $m$  with the exact solution (\ref{Rm}). Panel (b) compares the short-time tail (\ref{shorttime}) with numerical simulations.

\section{Coalescence and Decay}
\label{coalescence_and_death}
We now extend our analysis to a system undergoing both binary coalescence $2A\to A$ with rate $\lambda$ and linear decay $A\to 0$ with rate $\nu$. Although both these reactions reduce the number of particles in the system by $1$, they are  distinct from one another because the coalescence is a binary reaction, and its rate scales quadratically with the population size, rather than linearly as in the case of decay.  Thus, when the population size is large, the coalescence dominates the dynamics, and the decay term becomes relevant only when a few particles remain. Indeed, as we show below (see also the Appendix), the decay reaction does not contribute to $\mathcal{P}$ in the leading order. Its contribution appears only in the subleading order, and we will determine it in the Laplace space WKB formalism.

Rescaling time $t\to \lambda t$, and setting $r_n=\frac{1}{2}n(n-1)$ and $q_n=\mu n$, where $\mu=\nu/\lambda$, the forward and backward master equations read 
\begin{equation}
\label{master_coal}
\dot{P}_n(t)=r_{n+1}P_{n+1}(t)+q_{n+1}P_{n+1}(t) -(r_n+q_n)P_{n}(t)
\end{equation}
and
\begin{equation}
    \partial_T\mathcal{P}_m(T)=(r_m + q_m) \left[\mathcal{P}_{m-1}(T)-\mathcal{P}_m(T)\right].\label{back_master_coal}\,,
\end{equation}
respectively. Here too, before dealing
with ${\cal P}_m(T)$ we first compute the MTE, $\bar{T}_m$, which satisfies the recursive equation
\begin{equation}
\label{avtimeancoal}
\bar{T}_m=\bar{T}_{m-1} +1/(r_m+q_m)\,,
\end{equation}
with the boundary condition $\bar{T}_{m=0}=0$. The solution is \begin{equation}
\bar{T}=\frac{
2 \left[
H_m
+ \psi^{(0)}(2\mu)
- \psi^{(0)}(m+2\mu)
\right]
}{2\mu - 1},
\end{equation} 
where $H_m$ is the harmonic number and $\psi^0(z)=(d/dz)\ln{\Gamma{(z)}}$ is the digamma function.
In particular, in the limit of large $m$, we find
\begin{equation} \label{average_extinction_time_comp_death}
\bar{T}_{m\to \infty} = \frac{2\left[\psi^{(0)}(2\mu) + \gamma\right]}{2\mu - 1},
\end{equation}
where $\gamma\simeq 0.577$ is  Euler's constant.  (Note that no divergence occurs for $\mu=1/2$, for which $\bar{T}_{m\to \infty}=\pi^2/3$.)

\begin{figure}[ht]
\includegraphics[width=2.6 in,clip=]{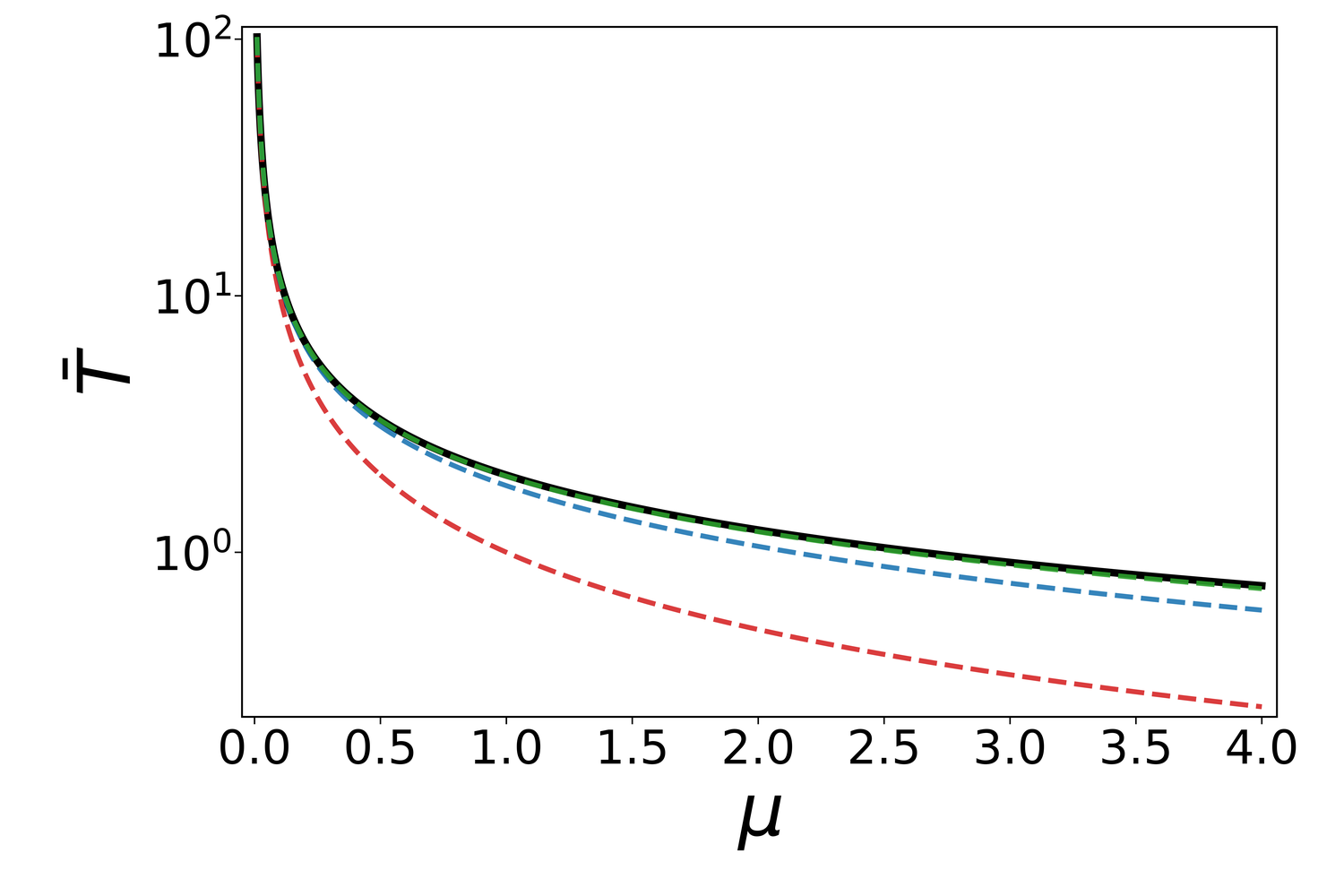}
\vspace{-5mm}\caption{ \justifying {The MTE $\bar{T}$ vs $\mu$ for the initial number of particles $m=1$ (dashed red line), $m=10$ (dashed blue line) and $m=100$ (dashed green line). Also shown is the MTE in the universal limit $m\to\infty$  (solid black line).}}
\label{fig3}
\end{figure}

For $\mu \to 0$ the MTE as described by Eq.~(\ref{average_extinction_time_comp_death}) diverges. This is to be expected since, in the absence of linear decay, the absorbing state at $n=0$ cannot be reached, and the system always enters the absorbing state $n=1$.
For $\mu=O(1)$ one obtains $\bar{T}_{m\to \infty}=O(1)$. As a result, the short-time tail of $\mathcal{P}(T)$ demands the strong inequality $T\ll1$.  Fig.~\ref{fig3} shows the MTE, as described by Eq.~(\ref{average_extinction_time_comp_death}), versus $\mu$ for several values of $m$. Noticeable is the convergence to the asymptotic value $\bar{T}_{m\to\infty}$ as $m$ is increased. Since $\bar{T}$ decreases monotonically with 
an increase of $\mu$, the applicability region of the short-time asymptotic of $\mathcal{P}(T)$ narrows. [Quantitatively, $\psi^{(0)}(z\gg 1)$ scales as $\ln z$, thus $\bar{T}_m$ decreases as $\ln \mu/\mu$ for large $\mu$.]

\vspace{-1cm}
\black{\subsection{Exact Result and Asymptotics}}

Applying the Laplace transform to the backward master equation~(\ref{back_master_coal}), we arrive at the equation
\begin{eqnarray} \label{comp_death}
        R(s,m)=\frac{(r_m+q_m) R(s,m-1)}{s+r_m+q_m}
\end{eqnarray}
with the boundary condition $R(s,m=0)=1$. This problem is exactly solvable, and the solution  is 
\begin{eqnarray}
\label{comp_death_R}
 R(s,m) = \frac{(1)_m (2 \mu)_m}{\left(\mu -\frac{\Psi}{2} +\frac{1}{2}\right){}_m \left(\mu +\frac{\Psi}{2}+\frac{1}{2}\right){}_m}   \,,
\end{eqnarray}
where for brevity we have denoted $\Psi=\sqrt{(1-2 \mu)^2-8s}$ and  used the Pochhammer symbol. Sending $m$ to infinity, we find the ``universal" asymptotic 
\begin{equation}
\label{exactRcoaldeath}
    \hspace{-3mm}R(s,m\to\infty)\!=\! \frac{\Gamma\left(\mu\!+\!\Psi/2\!+\!1/2\right)\Gamma\left(\mu\!-\!\Psi/2\!+\!1/2\right)}{\Gamma(2\mu)} \,.
\end{equation}
Thus, the $s\to\infty$ asymptotic of the exact result is
\begin{equation}\label{largescoal}
R(s\to\infty)=\frac{2\pi}{\Gamma (2 \mu )} (2s)^{\mu} e^{-\pi \sqrt{2s}}.
\end{equation}
Correspondingly, the short-time tail $\mathcal{P}(T\to 0)$ is given by the inverse Laplace transform of this asymptotic \black{(see, e.g. Ref. \cite{bateman1954tables})}: 
\begin{eqnarray} \label{short_tail_comp_death}
   \mathcal{P}(T\to 0) =\frac{\sqrt{2}}{\Gamma(2\mu)}\left(\frac{\pi}{T}\right)^{3/2 + 2\mu}
e^{-\frac{\pi^{2}}{2T}}.
\end{eqnarray}
As we can see, the decay reaction $A\to 0$ indeed contributes only to the subleading preexponential factor, while the leading-order exponential dependence  $\exp[-\pi^2/(2T)]$ is independent of $\mu$. Similarly to the annihilation example, this leading-order behavior is also captured by the leading-order time-dependent WKB calculation, see the Appendix. Yet, this calculation misses the large pre-exponential factor. 

\black{\subsection{Laplace-space WKB approximation}}

We now show how Eq. (\ref{largescoal}), and hence Eq. (\ref{short_tail_comp_death}),  can be reproduced  by our leading and subleading WKB formalism in the Laplace space, combined with an inner solution. 
As in the previous section, we make the WKB ansatz, see Eq.~(\ref{WKBansatz}), and assume that $S_0 \gg S_1$ and $s,m\gg 1$.  Applying the ansatz to 
Eq.~(\ref{comp_death}), we find  
\begin{eqnarray}
   S_0 '=\ln\left(1+\frac{2 s}{m^2}\right) , \quad S_1'= \frac{4 \mu s}{m^3 + 2ms}.
\end{eqnarray} 
Integrating the equations as before ($S_0$ from $0$ to $m$, and $S_1$ from  an a priori unknown cutoff $m_0$ to $m$) we find
\begin{eqnarray}\label{s0s1coal}
   && S_0 = 2\sqrt{2s}\arctan\left(\frac{m}{\sqrt{2s}}\right)+m\ln\left(1+\frac{2s}{m^2}\right), \nonumber \\
    &&S_1 = \mu\ln{\left(\frac{1+2s/m^2}{1+2s/m_0^2}\right)}.
\end{eqnarray}
As to be expected, the leading-order action $S_0(m)$ does not depend on $\mu$ and includes terms which are much larger than $1$, whereas  $S_1$ is ${\cal O}(1)$.

To calculate the cutoff $m_0$ we first need to find the inner solution. The latter solves a simplified variant of Eq.~(\ref{comp_death}):
\begin{equation}
    s R(s,m)=(r_m+q_m) R(s,m-1),
\end{equation}
where we have assumed that $R(s,m-1)\gg R(s,m)$ (to be checked a posteriori) and neglected terms which scale as $m^2 R(s,m)$ and $m R(s,m)$ compared to $s R(s,m)$. The boundary condition is  $R(s,m=0)=1$. The resulting inner solution is
\begin{equation}
     R_< =  \frac{m!}{(2s)^{m}} \frac{\Gamma(2\mu+m)}{\Gamma(2\mu)}.
\end{equation} 
The large-$m$ asymptotic of this solution is the following:
\begin{equation}
    R_<\simeq\frac{2\pi m^{2\mu}}{\Gamma(2\mu)} e^{-2m}\left(\frac{m^2}{2s}\right)^m\!.
\end{equation}
Now we match this asymptotic to the WKB solution~(\ref{WKBansatz}) in their joint  regime of validity $1\ll m\ll \sqrt{s}$. Here, the WKB solution~(\ref{WKBansatz}) reads
\begin{equation}
R_{WKB}= \left(\frac{m}{m_0}\right)^{2\mu}e^{-2m}\left(\frac{m^2}{2s}\right)^m\!.
\end{equation}
Matching the two solutions yields $m_0 = \left[\Gamma(2\mu)/2\pi\right]^{1/(2\mu)}$. Plugging this value of $m_0$ into $S_1$ in Eq.~(\ref{s0s1coal}) and taking the limit of $m\to\infty$ and $s\to\infty$, we recover Eq.~(\ref{largescoal}) in its entirety which leads us to Eq.~(\ref{short_tail_comp_death}).

\begin{figure}[ht]
\includegraphics[width=2.7 in,clip=]{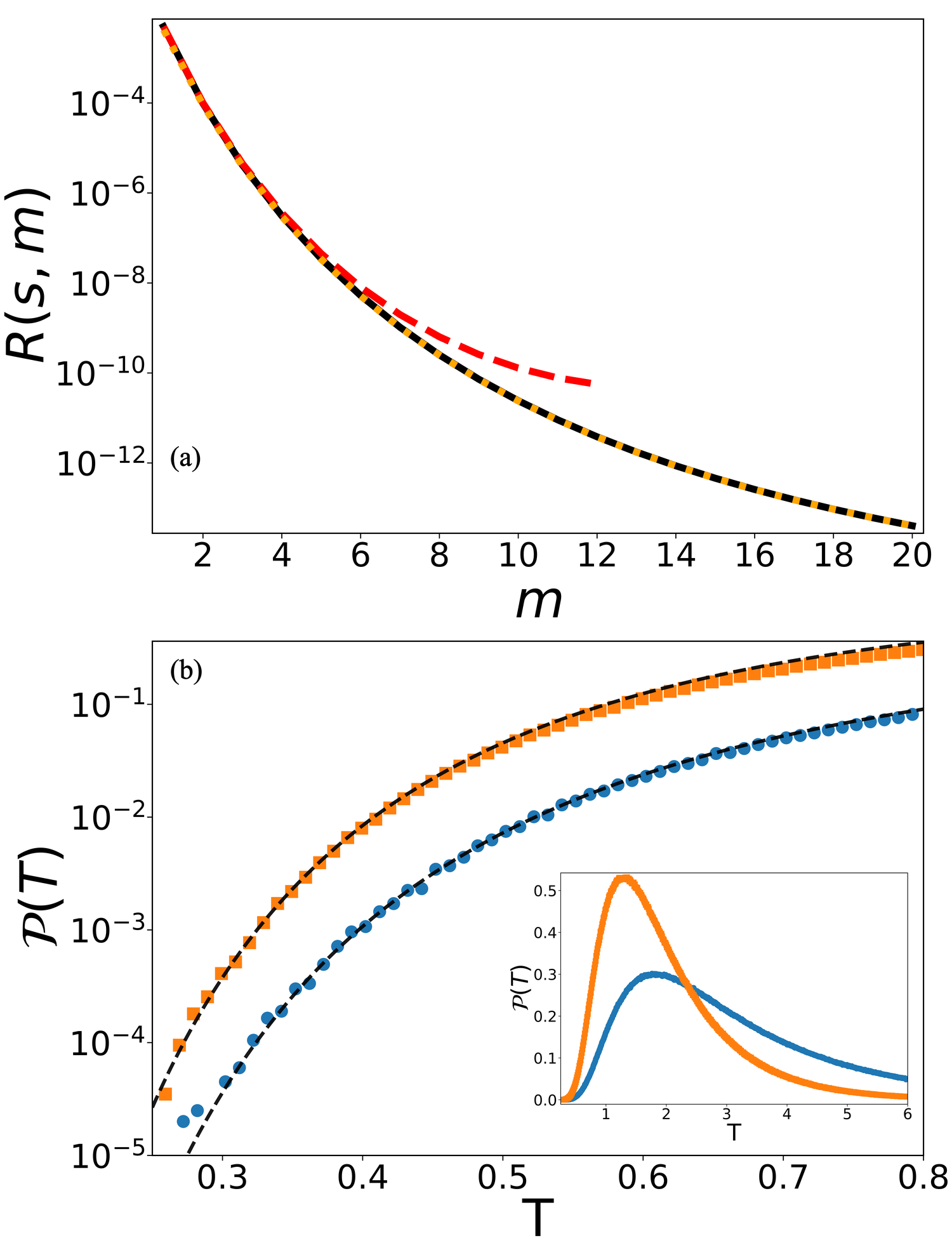}
\vspace{-5mm}\caption{ \justifying {(a) Solutions  of  Eq.~(\ref{comp_death}): Exact (solid black line), inner (red dashed line) and WKB (orange dotted line), with $s=10^2$.  (b) The probability distribution of extinction times $\mathcal{P}(T)$ for $\mu=1$ (orange squares) and $\mu=1/2$ (blue circles): the short-time tail~(\ref{short_tail_comp_death}) (line) versus numerical simulations (circles) with $2\times 10^7$ realizations. The inset shows simulation results for $\mathcal{P}(T)$  (in linear scale) over a longer time scale.}}
\label{fig4}
\end{figure}

Figure~\ref{fig4}(a) compares  the (leading and subleading) WKB solution and the inner solution for $R(s,m)$, as functions of $m$,  with the exact solution (\ref{comp_death_R}).  Panel (b) compares the short-time tail (\ref{short_tail_comp_death}) with numerical simulations. 

\section{Pair branching}
\label{superMalthus}
Sections \ref{Annihilation} and \ref{coalescence_and_death} dealt with population extinction. 
In this section we focus on another type of extreme first-passage events: a finite-time population blowup where, as we show, a similar large-deviation formalism applies. As an instructive example, we consider the pair branching reaction  $2A \to 3A$, which describes a super-Malthusian (that is, a faster-than-exponential) population growth and leads to a finite-time blowup of the number of particles \cite{M2025a}. Although the blowup time probability distribution $\mathcal{P}(T)$ in this case can be calculated exactly for any initial number of particles $m$ \cite{M2025a}, here we will circumvent the exact solution. Our objective is to show that the short-time tail of $\mathcal{P}(T)$ can be accurately determined by applying the leading and subleading Laplace-space WKB method combined with a non-WKB inner solution. The key difference between the blowup  and  extinction settings is that in the blowup case a natural initial number of particles is $O(1)$ which is beyond the applicability of the WKB approximation. For this reason the inner solution is necessary here not just for determining a constant multiplier $O(1)$, but for calculating the entire preexponential factor of the  $\mathcal{P}(T\to 0)$ tail.

The master equation for the double branching process reads
\begin{equation}\label{master}
 \!\dot{P}_n(t)\!=\! r_{n-1} P_{n-1}(t) - r_n P_{n}(t),
\end{equation}
where $r_n=n(n-1)/2$, and time is again rescaled by the rate constant.  The mean time to blowup when starting from $m$ particles is $\bar{T}_m= 2/(m-1)$ \cite{M2025a}. The short-time tail of $\mathcal{P}(T)$ correspond to stochastic realizations with a much shorter blowup time. 

\vspace{-0.5cm}
\black{\subsection{Exact Result and Asymptotics}}

The Laplace transformed backward equation satisfies:
\begin{equation}\label{Pieq}
r_m\left[R(s,m+1)-R(s,m)\right]=s R(s,m)\,,
\end{equation}
and it should be solved with the boundary condition
\begin{equation}
\label{BC2}
R(s,m\to+\infty)=1\,.
\end{equation}
This problem is exactly solvable for all $m>1$, and one obtains \cite{M2025a} 
\begin{equation}\label{exactbranching_m}
R(s,m) \!=\!\frac{\Gamma \left(m-\frac{1}{2}-\frac{1}{2} \sqrt{1-8 s}\right) \Gamma
   \left(m-\frac{1}{2}+\frac{1}{2} \sqrt{1-8 s}\right)}{\Gamma (m-1) \Gamma (m)}.
\end{equation}
In the benchmark case $m=2$ (when there are exactly two particles at $t=0$) the  solution gets simplified: 
\begin{equation}
\label{exactbranching}
 R(s,m=2)=\frac{2\pi s}{\cos\!\left(\frac{\pi}{2}\sqrt{1-8s}\right)}\,.
\end{equation}
The short-time tail of  $\mathcal{P}(T)$ is determined by the  $s\to \infty$ asymptotic of this $R(s,m)$, and one obtains~\cite{M2025a}
\begin{equation}\label{larges}
R(s\to \infty,2) \simeq  4\pi s \,e^{-\pi \sqrt{2s}}\,.
\end{equation}
The inverse Laplace transform of this expression gives the $\mathcal{P}(T\to 0)$ tail \cite{M2025a}
\begin{equation}\label{PsmallT}
  \mathcal{P}(T \to 0)  \simeq \frac{\sqrt{2}\, \pi ^{7/2}}{T^{7/2}}\, e^{-\frac{\pi ^2}{2 T}}\,,
\end{equation}
which exhibits an essential singularity at $T=0$ and a large power-law prefactor, as in the previous examples. 

\black{\subsection{Laplace-space WKB approximation}}

Now we will show how to reproduce  Eq.~(\ref{larges})~without using the exact solution. Assuming $s,m\gg 1$, and employing the same WKB ansatz for  $R(s,m)$ as before,  we obtain the following equations for $S_0$ and $S_ 1$: \black{
\begin{eqnarray}
   S_0 '=-\ln\left(1+\frac{2 s}{m^2}\right) , \quad S_1'= \frac{4 s}{m^3 + 2ms}.
\end{eqnarray}
}
As a result,
\begin{eqnarray}
\label{S0blow-up WKB}
  S_0(s,m) &=& \int_m^{\infty} \!\!\ln\left(1\!+\!\frac{2s}{z^2}\right) dz = \pi\sqrt{2s}-m \ln \left(m^2\!+\!2 s\right) \nonumber \\
           &-& \sqrt{8s} \arctan\left(\frac{m}{\sqrt{2s}}\right)+2 m \ln m,\nonumber\\
  S_1(s,m) &=&\ln \left(1+\frac{2s}{m^2}\right)\,,
\end{eqnarray}
where the two integration constants are chosen so as to obey the boundary condition (\ref{BC2}) which implies $S_0(m\to \infty) =S_1(m\to \infty) =0$.  The resulting WKB asymptotic $R_{\text{WKB}}(s\gg 1,m\gg 1)$ is given by Eq.~(\ref{WKBansatz}).
Now we return to Eq.~(\ref{Pieq}) and observe that, in the inner region $m\ll \sqrt{s}$ the second term on the l.h.s. can be neglected compared with the other two terms. The general solution of the resulting approximate recursive equation 
\begin{equation}\label{Pieqsimple}
r_m R(s,m+1)=s \Pi(s,m)
\end{equation}
is 
\begin{equation}\label{reducedsol}
R_<(s,m) = \frac{ 2^{m-2} K(s) s^{m-2}}{\Gamma (m-1) \Gamma (m)}\,,
\end{equation}
where $K(s)$ is an  arbitrary function to be determined.  An arbitrary function appears here in the region of not too large $m$ because the boundary condition  (\ref{BC2}) is imposed at $m\to \infty$, that is outside of the validity region of the simplified recursive solution (\ref{reducedsol}).

\begin{figure}[h]
\includegraphics[width=2.7 in,clip=]{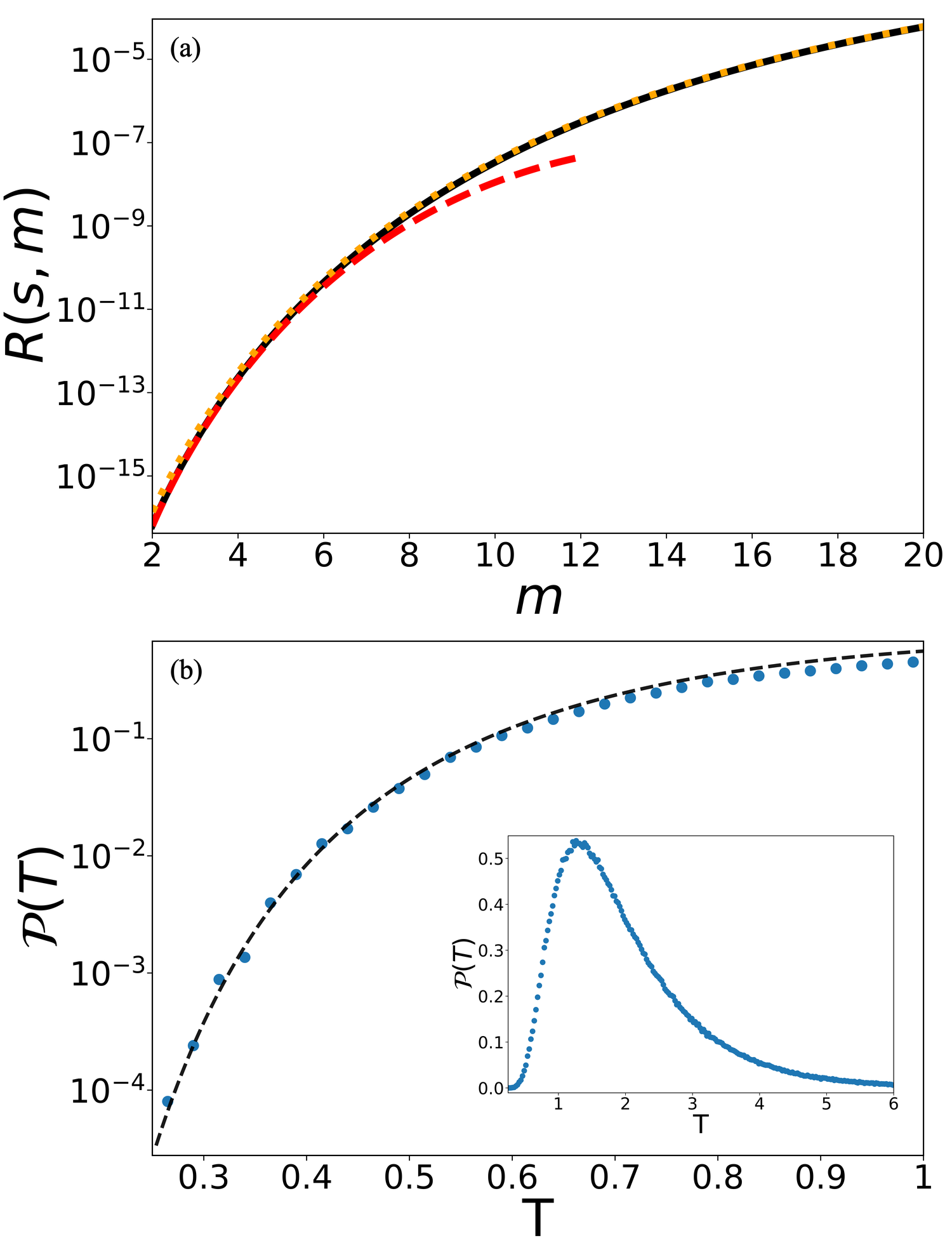}
\vspace{-4mm}\caption{ \justifying {(a) Solutions  of  Eq.~(\ref{Pieq}): exact (black solid line),  WKB (orange dotted line) and inner (red dashed line) for $s=10^2$. (b) The short-time tail (\ref{PsmallT}) of the blowup time distribution ${\cal P}(T)$ for the binary branching $2A \to 3A$  (dashed line) versus simulations (symbols) over $10^7$ realizations. The inset shows simulation results for ${\cal P}(T)$   (in linear scale) over a longer time scale.}}
\label{fig5}
\end{figure}

Now we can match the WKB solution (\ref{WKBansatz}) and the inner solution (\ref{reducedsol}) in their joint region of validity $1 \ll m \ll \sqrt{s}$. To this end we need the $m\ll \sqrt{s}$ asymptotic of $R_{\text{WKB}}$ and $m\gg 1$ asymptotic of $R_<(s,m)$:
\begin{equation}\label{asmore}
R_{\text{WKB}}(s,m\ll \sqrt{s}) \simeq  2^{m-1} e^{2 m-\pi \sqrt{2s}} \left(\frac{s}{m^2}\right)^{m-1}\,,
\end{equation}
and
\begin{equation}\label{asless}
R_<(s,m\gg 1) \simeq  \frac{2^{m-3} m^2  e^{2 m (1-\ln m)} K(s) s^{m-2}}{\pi }\,,
\end{equation}
respectively.  These two asymptotics coincide if and only if $ K(s) = 4 \pi s  e^{-\pi \sqrt{2s}}$. As a result, the inner solution 
(\ref{reducedsol}) becomes 
\begin{equation}\label{solreduced}
 R_<(s,m) = \frac{2^m \pi e^{-\pi\sqrt{2s}} s^{m-1}}{\Gamma (m-1) \Gamma (m)}\,.
\end{equation}
In particular,  $R_<(s,2)=K(s) = 4 \pi s  e^{-\pi \sqrt{2s}}$ which coincides with Eq.~(\ref{larges}) and, upon the inverse Laplace transform, leads to Eq.~(\ref{PsmallT}). 
As in Sec.~\ref{Annihilation}, a time-dependent WKB approximation can be used in order to reconstruct the short-time asymptotic of the extinction time distribution, and this has been done \cite{M2025a}. While successful in capturing the leading-order behavior $\exp[{-\pi^2/(2T)]}$, the time-dependent WKB approach overlooks the significant pre-exponential factor. 

In Fig.~\ref{fig5}(a) we compare the exact solution (\ref{exactbranching_m}) for $R(s,m)$ as a function of $m$ with the WKB solution (\ref{S0blow-up WKB}) and inner solution (\ref{solreduced}). Figure~\ref{fig5}(b)  compares the short-time tail $\mathcal{P}(T\to 0)$, see Eq.~(\ref{PsmallT}), with simulations. 

\vspace{0.5cm}
\section{Summary and Discussion}
\label{discussion}
\black{Intrinsic noise can dramatically accelerate population extinction or blowup. Here we focused on the situation} where the probability distribution $\mathcal{P}(T)$ of the extinction or blowup time $T$ exhibits an essential singularity as $T\to0$. 
We critically compared two methods of evaluation of the $\mathcal{P}(T \to 0)$ tail: a time-dependent WKB method, directly applied to the forward master equation of the process, and a (time-independent) WKB method applied to the Laplace-transformed backward master equation.   The time-dependent WKB approach is useful in predicting the optimal \black{---that is, the most likely---} path of the conditioned process, and in capturing a leading order behavior of the $\mathcal{P}(T \to 0)$ tail, including the essential singularity. It misses, however,  a pre-exponential factor of the $\mathcal{P}(T \to 0)$ tail \black{which can be $T$-dependent and therefore large}.  Moreover, this factor can depend on subdominant reaction channels, 
see  Sec.~\ref{coalescence_and_death}, and therefore carry important information about the underlying stochastic process.
To describe these, one can  go to a subleading order \black{of the time-dependent WKB method}, but such a calculation is difficult because of the intrinsic time-dependence of the problem. 

As we have shown here \black{(see also Ref. \cite{M2025b})}, the Laplace space WKB formalism, combined with the inner solution, avoids this difficulty and allows for a straightforward order-by-order perturbation theory, which fully recovers the short-time tail of the first passage time's probability distribution.  The solution involves matching between the WKB  and  inner asymptotics, which is possible due to the presence of a common region of validity where the two curves overlap.

\black{It is useful to put this formalism into a more general perspective. The application of backward master equations (or, for continuous Markov processes, of backward Fokker-Planck equations), in conjunction with the Laplace transform, to first-passage problems is by itself a standard technique, see e.g. Refs. \cite{Redner_2001,Krapivsky_Redner_Ben-Naim_2010}. Yet,  it has been mostly used previously in situations where \black{the Laplace transformed equation} is exactly solvable. A combination  of Laplace-space WKB approximation and the inner solution, based on the relevant small parameters, potentially extends the applicability range of the method.} 

\black{In their turn, different variants of the WKB approximation, in conjunction with inner-type solutions, were also used previously, but this was done in the context of exponentially slow extinctions or blowups of long-lived \emph{metastable} populations, see Refs.~\cite{OvaskainenM,Assaf_Meerson_2017} for reviews. In those problems the large parameter of the theory is the population size in the metastable state, and the formalism becomes effectively time-independent due to the exponential smallness of the relevant eigenvalue of the forward master equation operator. The examples we considered here are quite different both physically and mathematically, as they are intrinsically time-dependent, and the effective stationarity is achieved in the Laplace space.}

A future work can attempt extensions of this formalism to more complicated situations, including
rapid extinction or blowup of a population that would otherwise reach a long-lived metastable state.

\noindent
{\bf Acknowledgment}. B. M.  was supported by the Israel Science Foundation  (Grant No. 1579/25).

\bibliography{bibliography}

\appendix

\subsection*{Appendix: Time-Dependent WKB Theory for Sec.~\ref{coalescence_and_death} }\label{sec:time_WKB}

Here we employ the leading-order time-dependent WKB method, as applied directly to the master equation, in the case of binary coalescence and linear decay, $2A \to A$ and $A\to 0$, that we also consider in Sec.~\ref{coalescence_and_death}.

To compute the probability distribution of extinction times by the time-dependent WKB method in the coalescence decay problem, we recall the  master equation 
\begin{equation}
    \dot{P}_n(t)=r_{n+1}P_{n+1}(t)+q_{n+1}P_{n+1}(t) -(r_n+q_n)P_{n}(t).
\end{equation}
Assuming $n\gg 1$, using the ansatz $P_n(t) = \exp[-S(n,t)]$, and expanding $S(n+1,t)\simeq S(n,t)+S'(n,t)$, we arrive at the following Hamilton-Jacobi equation
\begin{equation}\label{HJeq_comp_death}
\partial_t S+H(n,\partial_n S) = 0\,,
\end{equation}
with the effective Hamiltonian\begin{equation}\label{H_com_death}
H(n,p) = \dfrac{n^2}{2}\left(e^{-p}-1\right).
\end{equation}
It is clear now that the decay reaction $A\to 0$  is irrelevant in the leading WKB order.  The Hamilton's equations are the following:
\begin{eqnarray}
  \dot{n} = - \dfrac{n^2}{2} e^{-p},\quad\quad \dot{p} = n \left(1-e^{-p}\right)\,.\label{neqcomp}
\end{eqnarray}
As the Hamiltonian is a constant of motion, we can denote $H(n,p)=E=const$, which allows us to express the momentum $p$ in terms of energy
\begin{equation}\label{pvsn_comp_death}
  p(n,E) = - \ln\left(1+2E/n^2\right)
\end{equation}
and obtain an autonomous equation for $\dot n$,
\begin{equation}\label{neq1_comp_death}
\dot{n} = -n^2/2-E.
\end{equation}
 Calculations similar to those shown in Sec.~\ref{Annihilation}, yield the following expression for the extinction time $T$ and the extinction action $S_0$
 \begin{eqnarray}
    &&T(E) = -\int_{\infty}^0 \frac{2dn}{n^2+2E}=\frac{2\pi}{\sqrt{8E}}, \\
    &&S_0=\int_{\infty}^0 p(n,E) dn - E T =  \sqrt{\frac{\pi^2E}{2}}=\frac{\pi^2}{2T}.
\end{eqnarray}
Here we have assumed that at $t=0$ there were infinitely many particles.
Recalling that the probability distribution of extinction times is $\mathcal{P}(T)=\exp{(-S_0(T))}$, we arrive at
\begin{equation}\label{P(T)lead_comp_death}
\mathcal{P}(T) \sim e^{-\pi^2/(2T)}.
\end{equation}
which coincides with the leading order result of Eq.~(\ref{short_tail_comp_death}). As in  Sec.~\ref{Annihilation}, the essential singularity  of ${\cal P}(T)$ is captured correctly. Nevertheless, this approximation misses the important pre-exponential factor of $T^{-(3/2+2\mu)}\gg 1$, see Eq.~(\ref{short_tail_comp_death}), and other $\mu$-dependent factors. This is to be expected, as the linear decay is irrelevant in the leading order.

For the optimal path itself, using Eq.~(\ref{neq1_comp_death}) with $E=\pi^2/(2T^2)$ we obtain
\begin{equation}\label{n(t)coal}
n(t) = \frac{\pi}{T}\cot \left(\frac{\pi  t}{2 T}\right)\,.
\end{equation}
Similarly to the purely annihilation problem, at $t\ll T$  it behaves as $n(t)\simeq 2/t$, which  coincides with the solution of the deterministic equation $\dot{n}(t)=-(1/2)n^2(t)$ for the initial condition $n(t=0) = \infty$.

\end{document}